\documentstyle[amssymb,12pt,epsfig,psfig]{article}
\makeatletter

\begin{document}

\title{ ON THE INFLUENCE OF ACOUSTIC WAVES ON COHERENT BREMSSTRAHLUNG IN CRYSTALS}
\author{A. A. Saharian\footnote{%
E-mail: saharyan@server.physdep.r.am}, A. R. Mkrtchyan,  V. V.
Parazian,
L. Sh. Grigoryan\\
{\it Institute of Applied Problems in Physics,}\\ {\it 25
Nersessian Str., 375014 Yerevan, Armenia}}

\maketitle

\begin{abstract}
We investigate the coherent bremsstrahlung by relativistic
electrons in a single crystal excited by hypersonic vibrations.
The formula for the corresponding differential cross-section is
derived in the case of a sinusoidal wave. The conditions are
specified under which the influence of the hypersound is
essential. The case is considered in detail when the electron
enters into the crystal at small angles with respect to a
crystallographic axis. It is shown that in dependence of the
parameters, the presence of hypersonic waves can either enhance or
reduce the bremsstrahlung cross-section.
\end{abstract}

\bigskip


\bigskip


\bigskip

\section{Introduction}

The processes converting the energy of relativistic electrons into
flows of electromagnetic radiation with the help of single
crystals are still of fundamental and practical interest in
high-energy physics. In crystals the cross-sections of the
high-energy electromagnetic processes can change essentially
compared with the corresponding quantities for a single atom (see,
for instance, Refs.
\cite{TerMik,Saen85,Shulga,Bazy87,Baie89,Rull98} and references
therein). The momentum transfer between a highly relativistic
interacting particle and the crystal can be small, especially
along the direction of particle motion. When this longitudinal
momentum transfer is small, the uncertainty principle dictates
that the interaction is spread out over a distance, known as the
formation length for radiation or, more generally, as the
coherence length. If the formation length exceeds the interatomic
spacing, the interference effects from all atoms within this
length are important and they can essentially affect the
corresponding cross-sections. From the point of view of
controlling the parameters of the high-energy electromagnetic
processes in a medium it is of interest to investigate the
influence of external fields, such as acoustic waves, temperature
gradient etc., on the corresponding characteristics. The
considerations of concrete processes, such as diffraction
radiation \cite{MkrtDR}, transition radiation \cite{Grigtrans},
parametric X-radiation \cite{Mkrt91}, channeling radiation
\cite{Mkrtch1}, electron-positron pair creation by high-energy
photons \cite{Mkrt03} have shown that the external fields can
essentially change the angular-frequency characteristics of the
radiation intensities.

The coherent bremsstrahlung of high-energy electrons moving in a
crystal is one of the most effective methods for producing of
quasimonochromatic gamma-quanta and has been intensively
investigated either theoretically and experimentally over the last
decade (see, for instance, Refs.
\cite{TerMik,Saen85,Shulga,Bazy87,Baie89,Rull98}). Such radiation
has a number of remarkable properties and at present it has found
many important applications. Among these is the generation of
intense positron beams. The basic source to creating positrons for
high-energy electron-positron colliders is the electron-positron
pair creation by hard bremsstrahlung photons produced when a
powerful electron beam hits an amorphous target. One possible
approach to increase the positron production efficiency is to use
a crystal target as a positron emitter (see Refs.
\cite{Artr03,Okun03} and references therein). When the crystal
axis is aligned to the incident beam direction, intense photons
are emitted through the coherent bremsstrahlung process and the
channeling radiation process. These photons are then converted to
electron-positron pairs in the same crystal, or in the amorphus
target behind the crystal.

The wide applications of the bremsstrahlung by relativistic
particles motivate the importance of investigations for various
mechanisms of controlling the radiation parameters. In the present
paper we investigate the influence of a hypersonic wave on the
coherent bremsstrahlung by relativistic electrons in a crystal. We
specify the conditions under which the external deformation field
changes noticeably the bremsstrahlung cross-section compared to
the case of an undeformed crystal and demonstrate the possibility
for the radiation yield enhancement. The plan of the paper is as
follows. In Sec. \ref{sec2:form} a formula is derived for the
coherent part of the bremsstrahlung by an electron in presence of
the sinusoidal deformation field generated by a hypersonic
vibrations. The analysis of the general formula and numerical
results in the special cases when the electron enters into the
crystal at small angles with respect to crystallographic axes or
planes are presented in Sec. \ref{sec3:an}. The main results are
summarized in Sec. \ref{sec4:conc}.

\section{Cross-section for the coherent bremsstrahlung} \label{sec2:form}

Let us consider the bremsstrahlung by a relativistic electron
moving in a single crystal excited by hypersonic vibrations. The
corresponding cross-section can be presented in the form (see, for
example, \cite{TerMik,Shulga})
\begin{equation}
\sigma ({\bf q})\equiv \frac{d^{4}\sigma }{d\omega d^{3}q}=
\left| \sum_{n}e^{i{\mathbf{qr}}_{n}}\right| ^{2}\sigma _{0}(%
{\bf q}),  \label{sig1}
\end{equation}
where ${\mathbf{q}}={\mathbf{p}}_1-{\mathbf{p}}_2-{\mathbf{k}}$ is
the momentum transferred to the crystal, $\sigma
_{0}({\mathbf{q}})$ is the cross-section on an individual atom,
${\mathbf{r}}_{n}$ are the positions of atoms in the crystal. Here
and below ${\mathbf{p}}_1$ and ${\mathbf{p}}_2$ are momenta of
particle in the initial and final states, $\omega $ and
${\mathbf{k}}$ are the frequency and wave vector for the radiated
photon (the system of units $\hslash =c=1$ is used). The
differential cross-section in a crystal, Eq. (\ref{sig1}), differs
from the cross-section on an isolated atom by the interference
factor which is responsible for coherent effects arising due to
periodical arrangement of the atoms in the crystal.

The positions of the atoms in a crystal can be presented as ${\bf
r}_{n}={\bf r}_{n0}+{\bf u}_{tn}$, where ${\bf u}_{tn}$ is the
displacement of atoms with respect to the equilibrium positions
${\mathbf{r}}_{n0}$ (by taking into account the crystal
deformation due to the hypersonic wave) due to the thermal
vibrations. After averaging on thermal fluctuations the
cross-section takes the standard form \cite{TerMik}
\begin{equation}
\sigma ({\bf q})=\left\{ N_{0}\left( 1-e^{-q^{2}\overline{u_{t}^{2}}}\right)
+e^{-q^{2}\overline{u_{t}^{2}}}\left| \sum_{n}e^{i{\bf qr}_{n0}}\right|
^{2}\right\} \sigma _{0}({\bf q}),  \label{sig2}
\end{equation}
where $\overline{u_{t}^{2}}$ is the temperature dependent
mean-squared amplitude of the thermal vibrations of atoms, $N_{0}$
is the number of atoms in the crystal,
$e^{-q^{2}\overline{u_{t}^{2}}}$ is the Debye-Waller factor. When
external influences are present the positions of atoms can be
written as
\begin{equation}
{\bf r}_{n0}={\bf r}_{ne}+{\bf u}_{n},  \label{rn0}
\end{equation}
with ${\bf r}_{ne}$ being the equilibrium positions of atoms in
the situation without deformation, ${\bf u}_{n}$ are the
displacements of atoms caused by the acoustic wave. We will
consider deformations with the sinusoidal structure
\begin{equation}
{\bf u}_{n}={\bf u}_{0}\sin \left( {\bf k}_{s}{\bf r}_{ne}+\varphi
_{0}\right) , \label{uacust}
\end{equation}
where ${\bf k}_{s}$ is the wave vector of the hypersonic wave.
Here the dependence of ${\bf u}_{n}$ on time (through the phase
$\varphi _{0}$) we can disregard, as for particle energies we are
interested in, the characteristic time for the change of
deformation field is much greater than the passage time of
particles through the crystal. For the deformation field given by
Eq. (\ref{uacust}) the sum over the atoms in Eq. (\ref{sig2}) can
be transformed into the form
\begin{equation}
\sum_{n}e^{i{\bf qr}_{n0}}=\sum_{m=-\infty }^{+\infty }J_{m}({\bf qu}%
_{0})e^{im\varphi _{0}}\sum_{n}e^{i{\bf q}_{m}{\bf r}_{ne}},\quad {\bf q}%
_{m}={\bf q}+m{\bf k}_{s},  \label{qm}
\end{equation}
where $J_{m}(x)$ is the Bessel function. For a lattice with a
complex cell the coordinates of the atoms can be written as ${\bf
r}_{ne}={\bf R}_{n}+{\bf \rho }_{j}$, with ${\bf R}_{n}$ being the
positions of the atoms for one of primitive lattices, and
${\mathbf{\rho }}_{j}$ are the equilibrium positions for other
atoms inside $n$-th elementary cell with respect to ${\bf R}_{n}$.
Now the square of the modulus for the sum (\ref{qm}) can be
presented as
\begin{equation}
\left| \sum_{n}e^{i{\bf qr}_{n0}}\right| ^{2}=\sum_{m,m^{\prime }=-\infty
}^{+\infty }J_{m}\left( {\bf qu}_{0}\right) J_{m^{\prime }}\left( {\bf qu}%
_{0}\right) e^{i(m-m^{\prime })\varphi _{0}}\sum_{n,n^{\prime }}e^{i{\bf q}%
_{m}{\bf R}_{n}}e^{-i{\bf q}_{m^{\prime }}{\bf R}_{n^{\prime }}}S({\bf q}%
_{m})S^{\ast }({\bf q}_{m^{\prime }}),  \label{SmSm}
\end{equation}
where $S({\mathbf{q}})=\sum_{j}e^{i{\mathbf{q}}{\mathbf{\rho
}}_{j}}$ is the structure factor of an elementary cell. For thick
crystals the sum over cells can be presented as a sum over the
reciprocal lattice:
\begin{equation}
\sum_{n}e^{i{\bf q}_{m}{\bf R}_{n}}=\frac{(2\pi )^{3}}{\Delta }\sum_{{\bf g}%
}\delta ({\bf q}_{m}-{\bf g}),  \label{sumdelta}
\end{equation}
where $\Delta $ is the unit cell volume, and ${\bf g}$ is the
reciprocal lattice vector. By taking into account the $\delta $ -
function, the quantity ${\bf q} _{m^{\prime }}$ can be written as
${\bf q}_{m^{\prime }}={\bf g}+(m^{\prime }-m){\bf k}_{s}$ and,
hence, we receive
\begin{equation}
\sum_{n^{\prime }}e^{-i{\bf q}_{m^{\prime }}{\bf R}_{n^{\prime
}}}=\sum_{n^{\prime }}e^{-i(m^{\prime }-m){\bf k}_{s}{\bf R}_{n^{\prime }}}=%
\frac{(2\pi )^{3}}{\Delta }\sum_{{\bf g}}\delta ((m^{\prime }-m)%
{\bf k}_{s}-{\bf g}).  \label{sumdelta1}
\end{equation}
As in the case of the electron-positron pair creation by
high-energy photons \cite{Mkrt03}, it can be seen that in the sum
over $m$ the main contribution comes from the terms for which
$m{\mathbf{k}}_s{\mathbf{u}}_0\lesssim
{\mathbf{g}}{\mathbf{u}}_0$, or equivalently $m\lesssim \lambda
_s/a$, where $\lambda _{s}=2\pi /k_{s}$ is the wavelength of the
external excitation, and $a$ is the lattice constant. Further,
under the condition $u_{0}/\lambda _{s}\ll 1$ the
contribution of the terms with $m\neq m^{\prime }$ in the sum (\ref{SmSm}%
) is small compared to the diagonal terms (see analogous
discussion in Ref. \cite{Mkrt03}). In the case $m=m^{\prime }$ the
sum in the left hand side of (\ref{sumdelta1}) is equal to the
number of cells, $N$, in a crystal and the square of the modulus
for the sum on the left of Eq. (\ref{SmSm}) can be written as
\begin{equation}
\left| \sum_{n}e^{i{\bf qr}_{n0}}\right| ^{2}=N\frac{(2\pi
)^{3}}{\Delta }\sum_{m=-\infty }^{+\infty
}J_{m}^{2}({\bf qu}_{0})\left| S({\bf q}%
_{m})\right| ^{2}\sum_{{\bf g}}\delta ({\bf q}_{m}-{\bf g}).  \label{SmSmnew}
\end{equation}
Note that in this case we have no dependence on the phase $\varphi _{0}$.

In formula (\ref{sig2}) the first two terms in figure braces do
not depend on the direction of the vector ${\bf q}$ and correspond
to the contribution of incoherent effects. The third summand
depends on the orientation of crystal axes with respect to the
vector ${\bf q}$ and determines the contribution of coherent
effects. The corresponding part of the cross-section is known as
an interference term. By taking into account the formula for
$\sigma _{0}({\bf q})$ (see, e.g., \cite{TerMik,Shulga}), in the
region $q\ll m_{e}$ for the values of the momentum transfer this
term can be written as
\begin{equation}
\sigma _{c}=\frac{e^{2}}{8\pi ^3 E_{1}^{2}}\frac{q_{\perp
}^{2}}{q_{\parallel }^{2}}\left| u_{{\bf q}}\right| ^{2}\left( 1+\frac{%
\omega \delta }{m_{e}^{2}}-\frac{2\delta }{q_{\parallel }}+\frac{2\delta ^{2}%
}{q_{\parallel }^{2}}\right) e^{-q^{2}\overline{u_{t}^{2}}%
}\left| \sum_{n}e^{i{\bf qr}_{n0}}\right| ^{2},  \label{sigcoh}
\end{equation}
where $E_1$ is the energy of the initial electron, ${\bf q}
_{\parallel }$ and ${\bf q}_{\perp }$ are the parallel and
perpendicular components of the vector ${\bf q}$ with respect to
the direction of the initial electron momentum ${\mathbf{p}}_1$,
$u_{ {\bf q}}$ is the Fourier-transform of the atomic potential,
$\delta =1/l_{c}$ is the minimum longitudinal momentum transfer,
and $l_{c}=2E_{1}E_{2}/(\omega m_{e}^{2})$ is the formation length
for the bremsstrahlung, with $E_2$ being the energy of the final
electron. Usually one writes the quantity $u_{{\bf q}}$ in the
form $4\pi Ze^2\left[ 1-F(q) \right] /q^{2}$, where $Z$ is the
number of electrons in an atom, and $F(q)$ is the atomic
form-factor. For the exponential screening of the atomic potential
one has $u_{{\bf q}}=4\pi Ze^2/(q^{2}+R^{-2})$, with $R$ being the
screening radius of the atom.

The total expression for the bremsstrahlung cross-section can be
presented in the form
\begin{equation}\label{gencross}
d\sigma =N_0(d\sigma _n+d\sigma _c),
\end{equation}
where $d\sigma _n$ and $d\sigma _c$ are the cross-sections for the
non-coherent and coherent bremsstrahlung in a crystal per single
atom. Using formulae (\ref{SmSmnew}) and (\ref{sigcoh}) and
integrating over ${\bf q}$, the cross-section for the coherent
part can be presented as
\begin{eqnarray}
\frac{d\sigma _{c}}{d\omega } &=&\frac{e^{2}N}{N_0 E_1^2 \Delta
}\sum_{m,{\bf g}}\frac{ g_{m\perp }^{2}}{g_{m\parallel
}^{2}}\left[ 1+\frac{\omega ^{2}}{2E_{1}E_{2}} -2\frac{\delta
}{g_{m\parallel }}\left( 1-\frac{\delta }{g_{m\parallel }}
\right) \right] \left| u_{{\bf g}_{m}}\right| ^{2}\times   \label{dsigpm} \\
&&\times J_{m}^{2}({\bf g}_{m}{\bf u}_{0})\left| S({\bf g})\right|
^{2}e^{-g_{m}^{2}\overline{u_{t}^{2}}},\quad {\bf g}_{m}={\bf
g}-m{\bf k}_{s},  \nonumber
\end{eqnarray}
where the summation goes under the constraint
\begin{equation}
g_{m\parallel }\geq \delta .  \label{sumcond}
\end{equation}
For the simplest crystal with one atom in the elementary cell one
has $N=N_{0}$ and $S({\mathbf{g}})=1$. Formula (\ref{dsigpm})
differs from the corresponding formula for the bremsstrahlung in
an undeformed crystal (corresponding to the summand with $m=0$ and
$u_{0}=0$, see, for instance, \cite{TerMik,Shulga}) by replacement
${\bf g}\rightarrow {\bf g}_{m}$, and additional summation over
$m$ with weights $J_{m}^{2}({\bf g}_{m}{\bf u}_{0})$. This
corresponds to the presence of an additional one dimensional
superlattice with period $\lambda _{s}$ and the reciprocal lattice
vector $m{\mathbf{k}}_s$, $m=0,\pm 1,\pm 2,\ldots $. Note that by
taking into account the $\delta $-function in Eq. (\ref{SmSmnew}),
the momentum conservation law can be written down as
\begin{equation}
{\bf p_1}={\bf p}_{2}+{\bf k}+{\bf g}-m{\bf k}_{s},
\label{conslaw}
\end{equation}
where $-m{\mathbf{k}}_s$ stands for the momentum transfer to the
external field.

\section{Discussion of the general formula and numerical results}
\label{sec3:an}

First of all let us specify the conditions under which the
influence of the external excitation on the bremsstrahlung
cross-section is noticeable. In formula (\ref{dsigpm}) the main
contribution comes from the terms with $g_{m\parallel }\sim \delta
$. It follows from here that the external excitation with a wave
vector ${\bf k}_{s}$ will influence on the process of the
bremsstrahlung if $mk_{s\parallel }\gtrsim \delta $. As a
consequence of the well-known properties of the Bessel function,
in the sum over $m$ the main contribution is due to the summands
with
\begin{equation}
m\lesssim {\bf g}_{m}{\bf u}_{0}\sim gu_{0}\sim \frac{2\pi
u_{0}}{a}. \label{mcond}
\end{equation}
From these relations it follows that it is necessary to take into
account the influence of external fields on the bremsstrahlung if
\begin{equation}
\frac{u_{0}}{\lambda _{s}}\gtrsim \frac{a}{(2\pi
)^{2}l_{c}}=\frac{am_{e}}{8\pi ^2}\frac{m_e\omega }{E_{1}E_{2}}.
\label{u0cond}
\end{equation}
It should be noted that at high energies $a/l_{c}\ll 1$ and
condition (\ref{u0cond}) does not contradict to the condition
$u_{0}/\lambda _{s}\ll 1$.

Let us consider the case of the simplest crystal with the
orthogonal lattice and one atom in the elementary cell assuming
that the electron enters into the crystal at small angle $\theta $
with respect to the crystallographic axis $z$. The corresponding
reciprocal lattice vector components are $g_i=2\pi n_i/a_i$,
$n_i=0,\pm 1,\pm 2,\ldots $, where $a_i$, $i=1,2,3$ are the
lattice constants in the corresponding directions. For the
longitudinal component we can write
\begin{equation}
g_{m\parallel }= g_{mz}\cos \theta +\left( g_{my}\cos \alpha
+g_{mx}\sin \alpha \right) \sin \theta ,  \label{gmpar}
\end{equation}
where $\alpha $ is the angle between the projection of the vector
${\mathbf{p}}_1$ on the plane $(x,y)$ and axis $y$. For small
angles $\theta $ the main contribution into the cross-section
comes from the summands with $g_{z}=0$ and we receive
\begin{equation}\label{crossc1}
\frac{d\sigma _c}{d\omega }\approx \frac{e^2}{E_1 ^2\Delta }\sum
_{m,g_x,g_y}\frac{g_{\perp }^2}{g_{m\parallel }^2}\left[
\frac{\omega ^2 }{2E_{1}E_{2}}+1-2\frac{\delta
}{g_{m\parallel}}\left( 1-\frac{\delta }{g_{m\parallel}} \right)
\right] |u_{g_{m}}|^2J_{m}^{2}({\mathbf{g}}_m{\mathbf{u}}_0),
\end{equation}
where $g_{\perp }^2=g_{x}^{2}+g_{y}^{2}$, and the summation goes
over the region $g_{m\parallel}\geq \delta $ with
\begin{equation}
g_{m\parallel }\approx -mk_{z}+\left( g_{y}\cos \alpha +g_{x}\sin
\alpha \right) \theta .  \label{gmparmain}
\end{equation}
Note that in the argument of the Bessel function
${\mathbf{g}}_m{\mathbf{u}}_0\approx {\mathbf{g}}_{\perp
}{\mathbf{u}}_0$. It follows from here that if the displacements
of the atoms in the acoustic wave are parallel to the axis $z$
then the main contribution into the cross-section is due to the
summand with $m=0$ and the influence of the acoustic wave is
small. The most promising case is the transversal acoustic wave
propagating along the $z$-direction. If the electron moves far
from the crystallographic plane (the angles $\alpha $ and $\pi
/2-\alpha $ are not small) the expression under the sum is a
smooth function on $g_x$ and $g_y$, and the summation over these
variables can be replaced by integration: $\sum _{g_x,g_y}\to
[a_1a_2/(2\pi )^2] \int dg_xdg_y $, and one receives
\begin{equation}\label{sumtoint}
\frac{d\sigma _c}{d\omega }\approx \frac{e^2}{4\pi ^2E_1 ^2a_1
}\sum _{m}\int dg_x dg_y \frac{g_{\perp }^2}{g_{m\parallel
}^2}\left[ \frac{\omega ^2 }{2E_{1}E_{2}}+1-2\frac{\delta
}{g_{m\parallel}}\left( 1-\frac{\delta }{g_{m\parallel}} \right)
\right] |u_{g_{m}}|^2J_{m}^{2}({\mathbf{g}}_m{\mathbf{u}}_0).
\end{equation}

\begin{figure}[tbph]
\begin{center}
\begin{tabular}{ccc}
\epsfig{figure=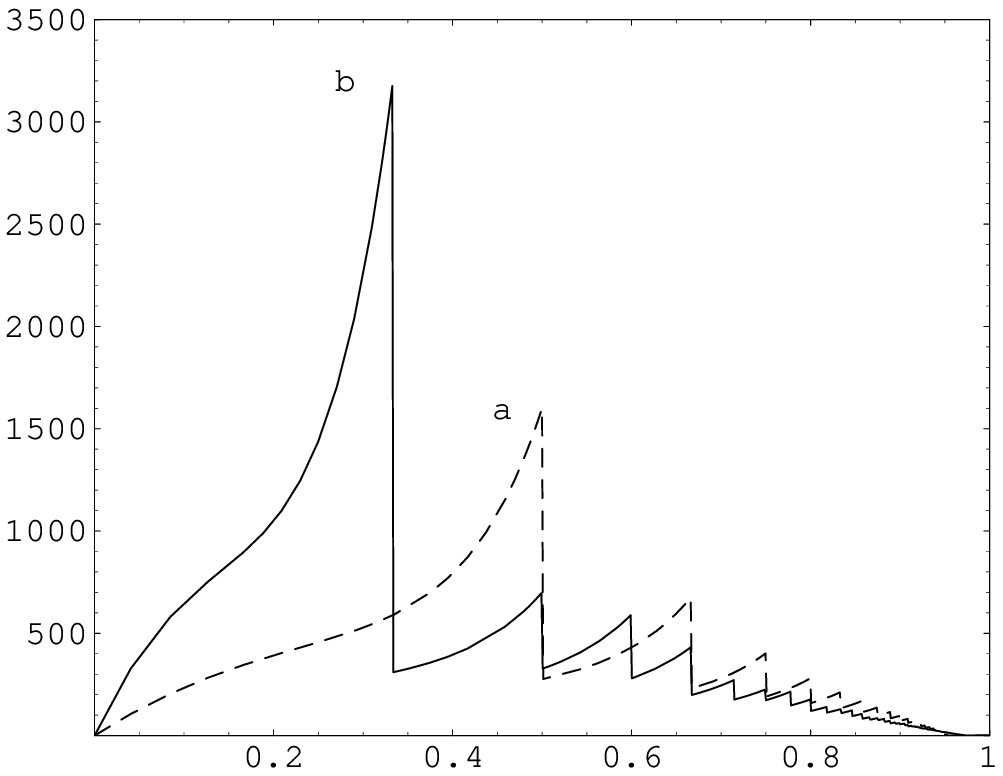,width=6cm,height=6cm} & \hspace*{0.5cm} & %
\epsfig{figure=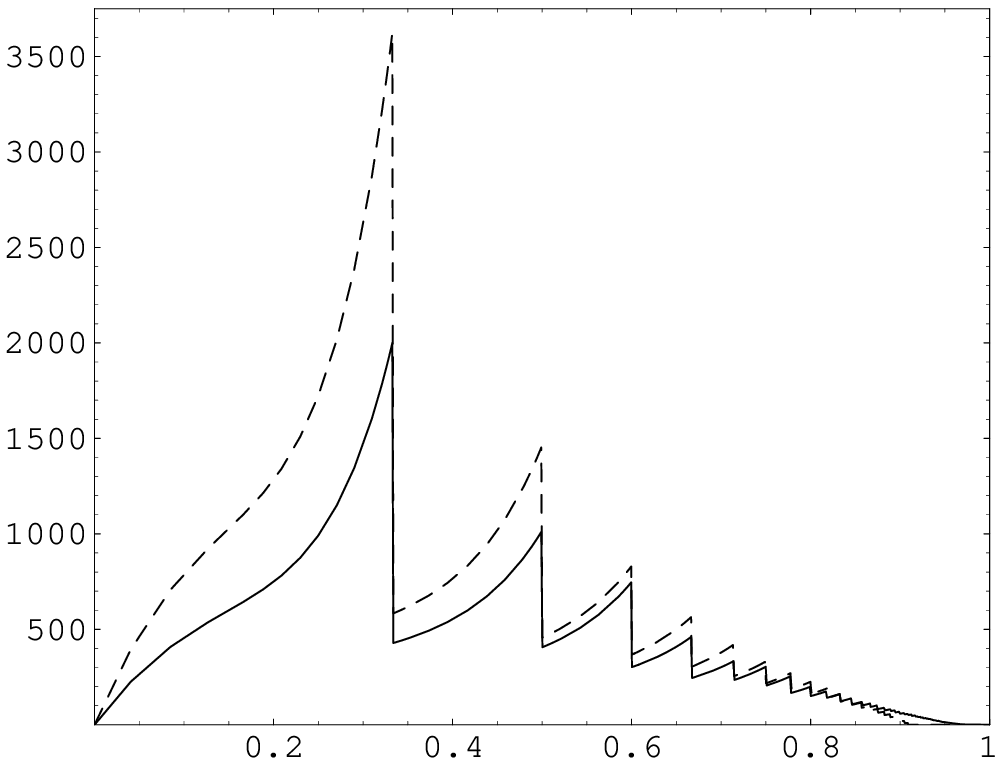,width=6cm,height=6cm}
\end{tabular}
\end{center}
\caption{Coherent bremsstrahlung cross-section, $(m_e^2\omega
/Z^2e^6)d\sigma _c/d\omega $, evaluated by formula
(\ref{farplane}), as a function of $\omega /E_{1}$ for $2\pi
u_0/a_2= 0$ (dashed curve), $2.2$ (full curve), $\theta =1$ mrad
(left panel) and $2\pi u_0/a_2= 0$ (dashed curve), $3$ (full
curve), $\theta =0.5$ mrad (right panel). The values for the other
parameters are as follows: $a_2/\protect\lambda _s=5\cdot 10^{-4}
$, $a_2/2\protect\pi R=1$, $m_e a_2 /(2\lambda _c E_1)=0.001$.}
\label{fig1tet}
\end{figure}

\begin{figure}[tbph]
\begin{center}
\begin{tabular}{c}
\psfig{figure=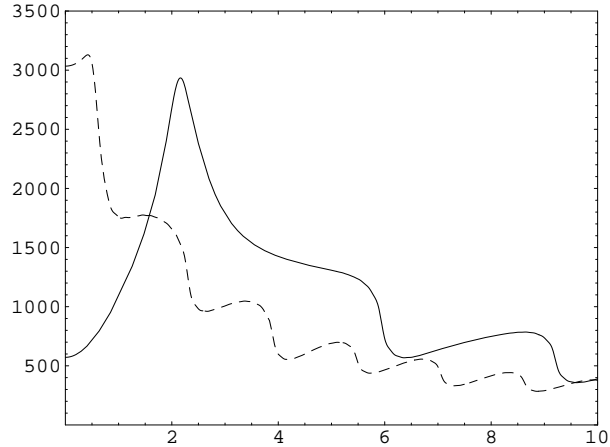,width=8cm,height=6cm}
\end{tabular}
\end{center}
\caption{Bremsstrahlung cross-section, $(m_e^2\omega
/Z^2e^6)d\sigma _c/d\omega $, evaluated by formula
(\ref{farplane}), as a function of $ 2\pi u_0/a_2$ for $\omega /
E_{1}=0.325$, $\theta =1$ mrad (full curve) and  $\omega /
E_{1}=0.25$, $\theta =0.2$ mrad (dashed curve). The values for the
other parameters are the same as in Fig.~\ref{fig1tet}.}
\label{fig2u0}
\end{figure}

We now assume that the electron enters into the crystal at small
angle $\theta $ with respect to the crystallographic axis $z$ and
near the crystallographic plane $(y,z)$ ($\alpha $ is small). In
this case with an increase of $\delta $ some sets of terms in the
sum will fall out. This can essentially change the cross-section.
Two cases have to be distinguished. Under the condition $\delta
\sim 2 \pi \theta /a_2$, in Eq. (\ref{crossc1}) for the
longitudinal component one has
\begin{equation}\label{condgy}
  g_{m\parallel }\approx
-mk_{z}+\theta g_{y}\geq \delta .
\end{equation}
This relation does not depend on the component $g_x$ and the
summation over this component can be replaced by integration $\sum
_{g_x}\to (a_1/2\pi )\int dg_x $. When ${\mathbf{u}}_0\parallel
{\mathbf{a}}_1$, for exponential screening the corresponding
integral is expressed in terms of the hypergeometric functions.
Here we will consider in detail the simpler case
${\mathbf{u}}_0\parallel {\mathbf{a}}_2$ when the variable $g_x$
does not enter in the argument of the Bessel function. For the
exponential screening after the elementary integration over $g_x$
we obtain
\begin{equation}\label{farplane}
  \frac{d\sigma _c}{d\omega }\approx \frac{4\pi ^2Z^2e^6}{E_1^2a_2a_3 }\sum
_{m,g_y}\frac{2g_{y}^2+R^{-2}}{g_{m\parallel}^2(g_y^2+R^{-2})^{3/2}}\left[
\frac{\omega ^2 }{2E_{1}E_{2}}+1-2\frac{\delta
}{g_{m\parallel}}\left( 1-\frac{\delta }{g_{m\parallel}} \right)
\right] J_{m}^{2}(g_yu_{0}),
\end{equation}
where the summation goes under the condition (\ref{condgy}). Note
that in this case the cross-section does not depend on the lattice
constant $a_1$. In Fig. \ref{fig1tet} we have plotted the coherent
part of the bremsstrahlung cross-section evaluated by formula
(\ref{farplane}) as a function on $\omega /E_1$ for $a_2=a_3$,
$a_2/\lambda _s=5\cdot 10^{-4}$, $a_2/2\pi R=1$, $m_e
a_2/(2\lambda _c E_1)=0.001$, where $\lambda _c$ is the Compton
wavelength of an electron. For the lattice constant $a_2\approx
1.89\cdot 10^{-8} $ cm the value of the last combination
corresponds to the electron energy $E_1 = 20$ GeV. On the left
panel of Fig. \ref{fig1tet} the graphics are plotted for $\theta
=1$ mrad, $2\pi u_0/a_2=0$ (dashed curve), $2.2$ (full curve) and
for the right panel we have chosen $\theta =0.5$ mrad, $2\pi
u_0/a_2=0$ (dashed curve), $3$ (full curve). The parameters are
taken in such a way to illustrate the fact that the hypersonic
wave can either enhance or reduce the cross-section. In Fig.
\ref{fig2u0} we give the dependence of cross-section
(\ref{farplane}) on the amplitude of the hypersound for $\omega /
E_{1}=0.325$, $\theta =1$ mrad (full curve) and $\omega /
E_{1}=0.25$, $\theta =0.2$ mrad (dashed curve).

Now we will assume that $\delta\sim 2 \pi \theta \alpha /a_1$. The
main contribution into the sum in Eq. (\ref{crossc1}) is due to
the terms with $g_y=0$ and we have two summations: over $m$ and
$n_1$, $g_x=2\pi n_1/a_1$. For the corresponding cross-section one
receives:
\begin{equation}\label{crossc2}
\frac{d\sigma _c}{d\omega }\approx \frac{e^2}{E_1^2\Delta }\sum
_{m,n_1}\frac{g_{x}^2}{g_{m\parallel}^2}\left[ \frac{\omega ^2
}{2E_{1}E_{2}}+1-2\frac{\delta }{g_{m\parallel}}\left(
1-\frac{\delta }{g_{m\parallel}} \right) \right]
|u_{g_{m}}|^2J_{m}^{2}(g_xu_{0x}),
\end{equation}
where
\begin{equation}\label{gmperpc2}
g_{m\parallel }\approx -mk_z+g_x\psi ,\quad \psi =\alpha \theta ,
\end{equation}
and summation goes over the values $m$ and $n_1$ satisfying the
condition
\begin{equation}
\left| n_1\psi -ma_{1}/\lambda _{s}\right| \geq
\frac{m_{e}^{2}a_{1}}{4\pi E_{1}E_{2}}. \label{sumcond2}
\end{equation}
In this case the most favorable conditions to have an influence on
the bremsstrahlung cross-section due to the hypersonic vibrations
are ${\bf u}_{0}\parallel {\bf a}_{1}$ (to have large values for
$m$) and ${\bf k}_{s}\parallel {\bf a}_{3}$ (to have large values
for $mk_{z}$).

\begin{figure}[tbph]
\begin{center}
\begin{tabular}{ccc}
\epsfig{figure=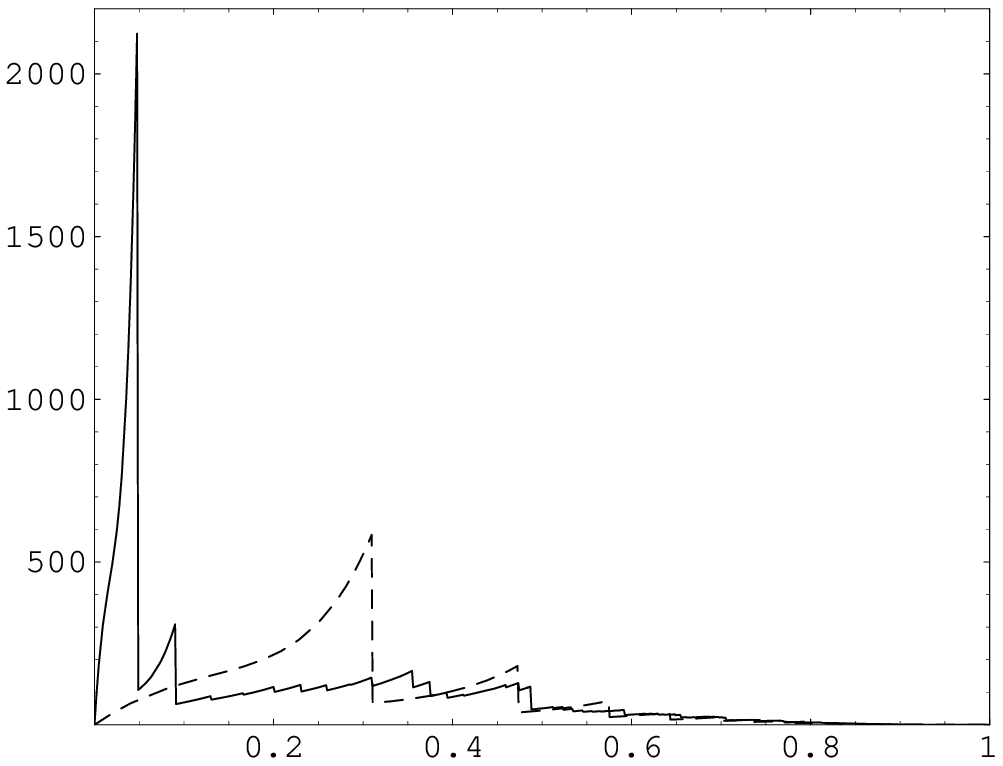,width=6cm,height=6cm} & \hspace*{0.5cm} & %
\epsfig{figure=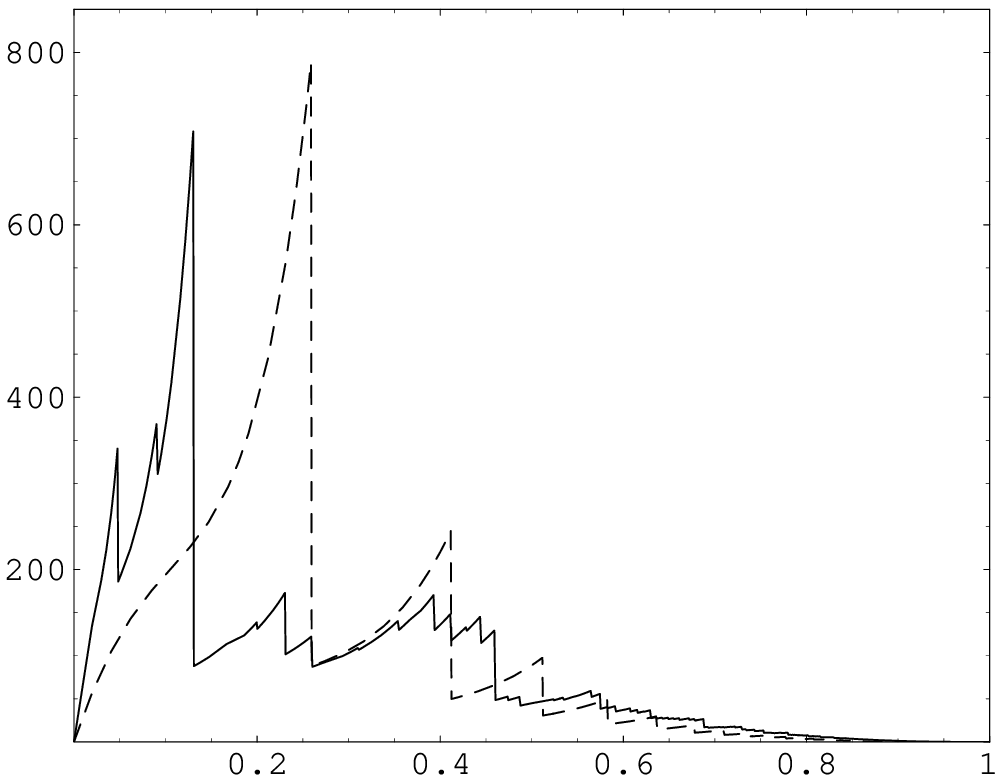,width=6cm,height=6cm}
\end{tabular}
\end{center}
\caption{Coherent bremsstrahlung cross-section, $(m_e^2\omega
/Z^2e^6)d\sigma _c/d\omega $, evaluated by formula
(\ref{crossc2}), as a function of $\omega /E_{1}/$ for $2\pi
u_0/a_1= 0$ (dashed curve), $2$ (full curve), $\psi =0.00045$
(left panel) and $2\pi u_0/a_1= 0$ (dashed curve), $2$ (full
curve), $\psi =0.00035$ (right panel). The values for the other
parameters are the same as in Fig. \ref{fig1tet} .}
\label{fig3psi}
\end{figure}

\begin{figure}[tbph]
\begin{center}
\begin{tabular}{c}
\psfig{figure=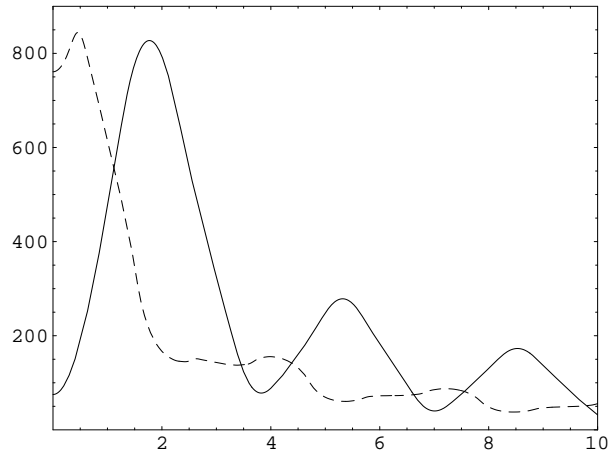,width=8cm,height=6cm}
\end{tabular}
\end{center}
\caption{Bremsstrahlung cross-section, $(m_e^2\omega
/Z^2e^6)d\sigma _c/d\omega $, evaluated by formula
(\ref{crossc2}), as a function of $ 2\pi u_0/a_1$ for $\omega /
E_{1}=0.1$, $\psi =0.00062$ (full curve) and  $\omega /
E_{1}=0.1$, $\psi =0.00017$ (dashed curve). The values for the
other parameters are the same as in Fig.~\ref{fig1tet}.}
\label{fig4u0}
\end{figure}

We have numerically evaluated the pair creation cross-section by
making use of formula (\ref{crossc2}) for various values of
parameters $\psi $, $u_0$, $\lambda _s$ and the energy of the
electron. As in the previous case, the corresponding results show
that, in dependence of these parameters, the external excitation
can either enhance or reduce the cross-section. As an illustration
in Fig. \ref{fig3psi} we have depicted the quantity $(m_e^2\omega
/Z^2e^6)d\sigma _c/d\omega $ as a function of $\omega /E_{1}$ in
the case of cubic lattice ($a_1=a_2=a_3$) and exponential
screening of the atomic potential for $u_{0}=0$ (dashed curve),
$2\pi u_{0}/a_{1}=2$ (full curve) and $\psi =0.00045$ (left panel,
both angles $\alpha $ and $\theta $ are measured in radians). For
the right panel $\psi =0.00035$, $u_{0}=0$ (dashed curve), $2\pi
u_{0}/a_{1}=2$ (full curve). The values for the other parameters
are the same as in Fig. \ref{fig1tet}. In Fig. \ref{fig4u0} we
have presented the cross-section evaluated by Eq. (\ref{crossc2})
as a function of $2\pi u_0/a_1$ for the photon energy
corresponding to $\omega /E_{1}=0.1$ and for $\psi =0.00062$ (full
curve), $\psi =0.00017$ (dashed curve). The values for the other
parameters are the same as in Fig. \ref{fig1tet}.

\section{Conclusion} \label{sec4:conc}

As a possible mechanism to control the spectral-angular
characteristics of the bremsstrahlung by relativistic electrons in
a crystal we have investigated the influence of the hypersonic
vibrations on the corresponding cross-section. If the
displacements of the atoms in the crystal under the influence of
hypersound have the form (\ref{uacust}), the coherent part of the
cross-section per single atom, averaged on thermal fluctuations,
is given by formula (\ref{dsigpm}). To compared with the
cross-section in an undeformed crystal this formula contains an
additional summation over the reciprocal lattice vector
$m{\mathbf{k}}_s$ of the one dimensional superlattice induced by
the hypersonic wave. The contribution for a given $m$ is weighted
by factor $J_m^2({\mathbf{g}}_m{\mathbf{u}}_0)$, where the vector
${\mathbf{g}}_m$ is defined as in Eq. (\ref{dsigpm}). We have
argued that the influence of the hypersound on the cross-section
can be remarkable under the condition (\ref{u0cond}). It should be
emphasized that for $u_0\gtrsim a$ this condition is less
restrictive than the naively expected one $l_c\gtrsim \lambda _s
$. In Sec. \ref{sec3:an} we have considered in detail the most
interesting case when the electron enters into the crystal at
small angle with respect to a crystallographic axis (axis $z$ in
our consideration). The main contribution into the coherent part
of the cross-section comes from the crystallographic planes,
parallel to the chosen axis. The behaviour of this cross-section
as a function on the photon energy essentially depends on the
angle between the projection of the electron momentum on the plane
$(x,y)$ and a crystallographic plane. If the electron moves far
from crystallographic planes, the summation over the perpendicular
components of the reciprocal lattice vector can be replaced by
integration and the coherent part of the cross-section is given by
formula (\ref{sumtoint}). When the electron enters into the
crystal near a crystallographic plane, two cases have to be
distinguished. For the first one $\theta \sim a_2/2\pi l_c$ and
the summation over $g_x$ can be replaced by integration. The
corresponding integral is easily evaluated in the case when the
displacements of atoms in the hypersonic wave are parallel to the
incidence plane, and the corresponding cross-section takes the
form (\ref{farplane}). The numerical results for this case are
presented in Figs. \ref{fig1tet} and \ref{fig2u0}. In the second
case $\psi =\alpha \theta \sim a_1/2\pi l_c$, and the main
contribution into the cross-section comes from the
crystallographic planes parallel to the incidence plane. The
corresponding formula has the form (\ref{crossc2}). The results of
the numerical evaluations for this case are depicted in Figs.
\ref{fig3psi} and \ref{fig4u0}. They show that in dependence of
the values for the parameters the presence of the hypersonic wave
can either enhance or reduce the cross-section.

\section*{Acknowledgment}

The work has been supported by Grant no. 1361 from Ministry of
Education and Science of the Republic of Armenia.

\end{document}